\documentclass[12pt]{article}
\usepackage[english]{babel}  
\usepackage[cp1251]{inputenc}
\usepackage[left=2cm, right=2cm, top=2cm]{geometry}

\usepackage{graphicx}
\usepackage{wrapfig}
\begin{document}

%
%
\centerline{\Large\bf Bistability and Synchronization in Coupled Maps Models}
\medskip
%
%
\centerline{ \textbf{Ricardo L\'{o}pez-Ruiz}  }
\medskip
%
%
\centerline{\it University of Zaragoza}
%
%
\centerline{\it rilopez@unizar.es}
\bigskip

Several coupled maps models are sketched and reviewed in this short communication.

First, a discrete logistic type model that was proposed for the symbiotic interaction 
of two species \cite {LR-1991,LR-2004}. The coupling between the species depends on the 
size of the other species $(x_n,y_n)$ at each time $n$ and of a constant $\lambda$ that 
it was called the {\it mutual benefit}:

\begin{eqnarray}
\label{eq1}
x_{n+1} & = & \lambda (3y_n+1)x_n(1-x_n) \\
y_{n+1} & = & \lambda (3x_n+1)y_n(1-y_n). \nonumber
\end{eqnarray} 

\begin{wrapfigure}{r}{0.4\textwidth}
\centering
\includegraphics[bb=0 0 400 400,scale=0.4]{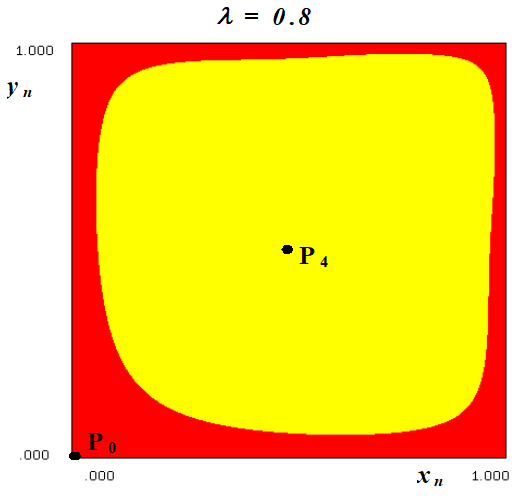}
\caption{\label{fig1}\scriptsize \textit{Bistability between the fixed populations 
$P_0$ and $P_4$ and their basins of attraction for the map~(\ref{eq1}) when $\lambda=0.8$.}}
\end{wrapfigure}

If the species are isolated, we recover the logistic model
for each of them. For $\lambda$ values where one isolated 
species would extinguish, the coexistence with the other species allows
them to have an alternative state to survive, supporting the idea that 
the symbiosis between both species makes possible the {\it bistability} (Fig.~\ref{fig1}).
For increasing values of $\lambda$ different dynamical regimes are obtained, 
from periodic oscillations of the populations, through
quasiperiodicity up to a chaotic regime before the final collapse.

\begin{wrapfigure}{r}{0.4\textwidth}
\centering
\includegraphics[bb=0 0 400 400,scale=0.4]{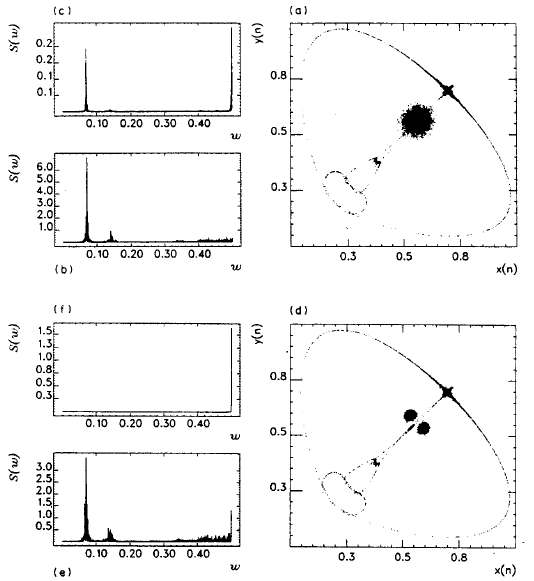}
\caption{\label{fig2}\scriptsize \textit{$\bf{(a-c)}$ Iterates and Fourier spectra of the globally coupled map~(\ref{eq2}) 
for $\lambda=1.07$ when $\epsilon=0.01$ and  $N=100$: (a) Iterates of a single element (the exterior attractor) 
and of the average (the central spot), (b) temporal Fourier spectrum of a single unit (arbitrary units),
(c) temporal Fourier spectrum for the average (mean field). $\bf{(d-f)}$ The same as (a-c) when $N=1000$. 
The synchronization effect mediated by the increasing size $N$ is observed with the appearance 
of a period-$2$ oscillation (at $w=1/2$) in the mean field, and by extension in the single element.}}
\end{wrapfigure}

Second, if many of these symbiotic species, $i=1,\cdots, N$, evolve together $(x_n^i,y_n^i)$
under the coupling action of a global mean field, $X_n^N$ and $Y_n^N$, with strength $\epsilon$,  

\begin{eqnarray}
\label{eq2}
x_{n+1}^i & = & f_{\lambda,N}(x_n^i, y_n^i) \\
y_{n+1}^i & = & f_{\lambda,N}(y_n^i, x_n^i) \nonumber
\end{eqnarray} 

\noindent with 

\[
f_{\lambda,N}(x_n^i, y_n^i) =  
          \lambda \left[3\left(\frac{y_n^i+\epsilon Y_n^N}{1+\epsilon}\right)+1\right] x_n^i(1-x_n^i) 
\]

\noindent and
$X_n^N= \frac{1}{N}\sum_{i=1}^{N} x_n^i  \;\; , \;\; Y_n^N= \frac{1}{N}\sum_{i=1}^{N} y_n^i$,
then the synchronization phenomenon of many oscillators appears, not only under the effect 
of a strong coupling $\epsilon$ if not also under the effect of an increasing size $N$ of the system. 
This is one of the first models appeared in the literature where the {\it synchronization} of many chaotic 
oscillators was reported \cite{LR-1991}. Thus, it can be easily seen that when $\epsilon=0$ the 
species (maps) are uncoupled and they evolve freely but when $\epsilon\rightarrow\infty$
the maps are totally synchronized and they evolve with the same dynamics all of them, in 
a periodic, quasiperiodic or chaotic state depending on the mutual benefit $\lambda$. 
But it is still more surprising the effect provoked by the size of the system consisting in 
that when $N$ increases there is an induced synchronization in this many agents model, 
such as it was also reported in Ref.~\cite{LR-1991} and it is sketched in Fig.~\ref{fig2}.

Third, if the species $x_i$ interact in a symbiotic way with the local mean field $X_n^i$ coming from 
the nearest neighboring species then we can implement this model in different kind of networks. 
This implementation was done in Ref.~\cite{LR-2007} interpreting this model in a neuronal context 
as a naive brain model where the bistability of the global system mimics the sleeping and 
the awaking states of a primitive brain. These are the two global synchronized states, 
the turned off ($x_{\theta}$) and the turned on ($x_+$) states of the network in question, 
that in the context of species would mean the extinction or the coexistence of the species in the ecosystem.
This model is

\begin{equation}
x^i_{n+1} = \bar p_i\;x^i_n(1-x^i_n)
\label{eq3}
\end{equation}

\noindent where $\bar p_i$ is the symbiotic parameter that depends on the local mean value, $X_n^i$, generated 
by the neighboring species, $\bar p_i = p\;(3X_n^i+1)$, with $X_n^i={1\over N_i}\sum_{j=1}^{N_i}x_n^j$.
$N_i$ is the number of neighbors of the $ith$ species (node $i$ of the network), and $p$ is the
actual version of the mutual benefit $\lambda$ of the model~(\ref{eq1}).
The synchronized states of system~(\ref{eq3}) can be analytically found.
The solutions are $x_\theta=0$ and $x_+\neq 0$. The first state $x_\theta$ is stable for $0<p<1$
and the state $x_+$ is stable for $p>0.75$. Therefore the bistability between both states,
$(x_n^i = x_\theta, x_n^i = x_+), \forall i$,
is possible for $p>p_0=0.75$ in the case of many interacting species on a network.
Let us observe at this point that the non-null state $x_+$ appears in the system as
a kind of {\it explosive synchronization} mediated by a global saddle-node bifurcation in the network that takes place
exactly at $p=p_0$, but in this case the appearance of the new synchronized state $x_+$ coexists with another non-noisy
synchronized state $x_\theta$. Also, different routes to enhance or switch on these both states and other
similar multi-dimensional models where reported at that time in Refs.~\cite{LR-2007,LR-2012}. 

Finally, some new results concerning this last model embedded in different topologies will be presented \cite{LR-2020}. 

\begin{thebibliography}{10}
\small

\bibitem{LR-1991} 
R.~L\'{o}pez-Ruiz, C.~P\'{e}rez-Garc\'{i}a,
      Dynamics of maps with a global multiplicative coupling, 
      Chaos, Solitons and Fractals 1 (1991) 511--528.
      
\bibitem{LR-2004} 
R.~L\'{o}pez-Ruiz, D.~Fournier-Prunaret ,
      Complex behaviour in a discrete logistic model for the symbiotic interaction of two species, 
      Math. Biosci. Eng. 1 (2004) 307--324.
     	  
\bibitem{LR-2007} 
R.~L\'{o}pez-Ruiz, Y.~Moreno, A.P.~Fern\'{a}ndez, S.~Boccaletti, D.U.~Hwang,
      Awaking and sleeping of a complex network,
      arXi:nlin/0406053 (2004); Neural Networks 20 (2007) 102--108.

\bibitem{LR-2012} 
R.~L\'{o}pez-Ruiz, D.~Fournier-Prunaret ,
      The bistable brain: A neuronal model with symbiotic interactions,
      Ch. 10 in {\it Symbiosis: Evolution, Biology and Ecological Effects},
      A.F. Camisao \& C.C. Pedroso (Ed.), Nova Books (2012) 235--254.

\bibitem{LR-2020} 
R.~L\'{o}pez-Ruiz, M.~As\'{i}s-C\'{a}novas ,
      to be published.



\end{thebibliography}
\end{document}